\documentclass[sigconf]{acmart}

\usepackage{microtype}
\usepackage{graphicx}
\usepackage{subfigure}
\usepackage{booktabs} 
\usepackage{amsmath}
\usepackage{amsthm}
\usepackage{enumitem}
\usepackage{algorithmic}
\usepackage{algorithm}

\AtBeginDocument{%
  \providecommand\BibTeX{{%
    \normalfont B\kern-0.5em{\scshape i\kern-0.25em b}\kern-0.8em\TeX}}}

\copyrightyear{2020}
\acmYear{2020}
\setcopyright{acmcopyright}
\acmConference[KDD '20] {26th ACM SIGKDD Conference on Knowledge Discovery and Data Mining}{August 23--27, 2020}{Virtual Event, USA}
\acmBooktitle{26th ACM SIGKDD Conference on Knowledge Discovery and Data Mining (KDD '20), August 23--27, 2020, Virtual Event, USA}
\acmPrice{15.00}
\acmDOI{10.1145/3394486.3403321}
\acmISBN{978-1-4503-7998-4/20/08}

\begin{document}
\fancyhead{}
\title{Faster Secure Data Mining via Distributed Homomorphic Encryption}

\author{Junyi Li}
\affiliation{
  \institution{Electrical and Computer Engineering \\ University of Pittsburgh}}
\email{junyili.ai@gmail.com}
\author{Heng Huang}
\affiliation{
  \institution{JD Finance America Corporation}
  \institution{University of Pittsburgh}
}
\email{henghuanghh@gmail.com}

\renewcommand{\shortauthors}{Junyi Li and Heng Huang}

\begin{abstract}
Due to the rising privacy demand in data mining, Homomorphic Encryption (HE) is receiving more and more attention recently for its capability to do computations over the encrypted field. 
By using the HE technique, it is possible to securely outsource model learning to the not fully trustful but powerful public cloud computing environments.
However, HE-based training scales badly because of the high computation complexity. It is still an open problem whether it is possible to apply HE to large-scale problems.
In this paper, we propose a novel general distributed HE-based data mining framework towards one step of solving the scaling problem. The main idea of our approach is to use the slightly more communication overhead in exchange of shallower computational circuit in HE, so as to reduce the overall complexity. We verify the efficiency and effectiveness of our new framework by testing over various data mining algorithms and benchmark data-sets. For example, we successfully train a logistic regression model to recognize the digit 3 and 8 within around 5 minutes, while a centralized counterpart needs almost 2 hours.
\end{abstract}




\maketitle

\section{Introduction}
The past decade has witnessed a surge of revolutionary breakthroughs in high technology, accompanied with an avalanche of the large-scale data in many fields. 
In industry, the data is no longer hosted on a single server or cluster and more computational resources are required to analyze these data. The cloud computing and distributed machine learning techniques have emerged to be the ideal solutions for big and cost-effective data storage and analysis. However, at the same time, it has also raised security and privacy concerns for using sensitive user data in public cloud computing, especially in the financial and medical fields. As a result, states across the world enact laws to protect the privacy of user data. For example, the famous General Data Protection Regulation (GDPR) \cite{regulation2016regulation} passed by European Union in 2018 sets up strict rules for companies to use personal data. Traditional encryption techniques mainly address the security and privacy issues in the data storage and transfer tasks, but not the data analysis task. This poses great challenges to the machine learning and data mining researchers, \emph{i.e.} how can we securely learn the machine learning and data mining models in the public cloud computing and distributed machine learning environments, which may not be fully trustful. 

Existing approaches proposed can be roughly divided into four main categories, \emph{i.e.} Differential Privacy (DP) \cite{dwork2008differential, dwork2014algorithmic}, Secure Multi-Party Computation (MPC) \cite{yao1986generate}, Homomorphic Encryption (HE) \cite{gentry2009fully, van2010fully} and secure Enclave \cite{mckeen2013innovative, champagne2010scalable}. Every approach has its advantages and drawbacks: 
\begin{itemize}
    \item \textbf{Differential privacy (DP)} has strong theoretical guarantees to preserve privacy but leads to performance degradation which is not acceptable in many cases, what's more, privacy parameters choice in DP such as $\epsilon$ is still a mystery; 
    \item \textbf{Secure Multi-Party Computation (MPC)} enables multiple parties to securely evaluate a function without revealing their own input, but it heavily relies on communication and is usually not trivial to formulate a model under MPC framework; 
    \item \textbf{Homomorphic Encryption (HE)} is well known for its ability to do computations in the encrypted field, but it has notoriously high computation overhead; 
    \item \textbf{Secure Enclave} outsources the model training to a trusted third party. Obviously, its security guarantee is totally relied on the design of this third party system. There have been many attacks targeted at such kind of systems like the famous side channel attack. 
\end{itemize}
To sum up, it is still challenging to deploy a large-scale secure data mining system based on the current status of these techniques. In this paper, we want to take one step further to the building of a practical secure data mining system by focusing on decreasing the time complexity of the HE based data mining systems.

More precisely, we propose a novel distributed secure data mining framework which achieves fast model learning over Homomorphic Encryption. The design of our framework is based on identifying a key feature of the Homomorphic Encryption: HE actually performs approximate computation and the noise accumulated in the ciphertext after each operation. Naturally, plaintext can not be correctly recovered if the noise grows too much. As a result, a technique called bootstrap has been invented by the cryptographic researchers to control the noise. However, bootstrap is a major source of time complexity. It is  both extremely expensive (at least 10$\times$ slower compared to other basic operations) and leads to higher complexity of other operations. Naturally, we can save a lot of computation if the bootstrap is not performed. In this paper, we will show that this is possible for data mining models especially under the distributed learning scenario. Actually, data mining models are relatively insensitive to the noise. As shown by \citet{zhu2016trained}, it is possible to compress the model parameters as low as 1 bit without losing much performance.  More importantly, the depth of the computational circuit in distributed learning is bounded as the worker will refresh its parameter every a few iterations. 
Our framework is based on this simple observation and we will show later in the experimental section that our framework achieves at least 10$\times$ times of training speed boost while keeping the same model performance. As a simple example, our framework trains a logistic regression model over MNIST data-set to the accuracy of 96.5\% within around 5 minutes in contrast with nearly 2 hours training time of a traditional HE based learning framework. Finally, note that our system protects the data privacy in that the worker has access only to the encrypted data and have no idea about the original data.

\noindent\textbf{Contribution.} We summarize the main contributions of this paper as follows:
\begin{itemize}
    \item We demonstrate the possibility of doing bootstrap-free homomorphic encryption based data mining under the distributed machine learning setting for the first time;
    \item We show our new secure data mining framework can achieve at least 10$\times$ training time saving compared to the centralized HE-based data mining system while maintaining the same model prediction performance.
\end{itemize}

\noindent\textbf{Organization.} The remaining part of this paper is organized as follows: In Section~\ref{sec:related-work}, we introduce the related work, especially the recent development of homomorphic encryption and its application to the data mining; In Section~\ref{sec:preliminary}, we present some preliminary results as the basis of our framework; In Section~\ref{sec:framework-design}, we formally introduce our framework and give a practical algorithm that can be used for training; In Section~\ref{sec:experimental-results}, we present the experimental results and make discussions; In Section~\ref{sec:concluson}, we summarize this paper and discuss the future work.

\section{Related Work}
\label{sec:related-work}
HE supports arithmetic computation over the encrypted field. \lq~Homomorphic~\rq\ here means that the encryption function and decryption function are homomorphism between plaintext and ciphertext. More precisely, suppose $z_i$ denotes the plaintext, $s_i$ denotes the ciphertext and $\pi$ and $\pi^{-1}$ denote the encryption and decryption function respectively. If we have:
\begin{equation}
\begin{split}
    s_1 &= \pi(z_1) \\
    s_2 &= \pi(z_2)
\end{split}
\end{equation}
and:
\begin{equation}
\begin{split}
    z_0 &= z_1\;\otimes\;z_2 \\
    s_0 &= s_1\;\tilde{\otimes}\;s_2 
\end{split}
\end{equation}
Then we have $s_0 = \pi(z_0)$. $\otimes$ denote any arithmetic computation \emph{e.g.} addition and multiplication, while $\tilde{\otimes}$ denotes the encrypted version of the corresponding computation.

HE, which is viewed as the Holy Grail of cryptography \cite{tourky2016homomorphic}, dates back to the 70s when RSA technique was proposed. Different types of HE schemas are proposed since then. \citet{paillier1999public} proposes a framework that only supports homomorphic addition. Frameworks like Paillier's which only support certain HE operations are referred as partially HE later, while more powerful frameworks  like \cite{brakerski2014leveled} which support both addition and multiplication are called somewhat (leveled) HE. They are not fully-HE framework as they constrain the maximum depth of the computational circuit. The first real fully-HE (FHE) framework which supports the evaluation of arbitrary computational circuit is proposed by \citet{gentry2009fully} in 2009. A key innovation of this framework is the use of bootstrap technique to control the noise of the ciphertext. Actually, the core idea of bootstrap is just to re-encrypt the ciphertext but in a homomorphic way, \emph{i.e.} Define a computational circuit in the encryption field that is homomorphic to the operation of decryption and encryption and then evaluate this circuit. The initial design proposed by \citet{gentry2009fully} is extremely slow (several orders slower than other basic homomorphic operations such as addition and multiplication). An important research topic \cite{brakerski2014efficient, brakerski2011fully, van2010fully} afterwards is to improve the practicability of the fully HE proposed by \citet{gentry2009fully}. A good review of the recent development of HE can be found in~\citet{challa2020homomorphic}. Specially, our framework is built upon a recently proposed FHE framework named Homomorphic Encryption for Arithmetic of Approximate Numbers (HEAAN). This framework has two main features. Firstly, it supports homomorphic rounding operation. As a result, the magnitude of the ciphertext modulus is linear with respect to the depth of the computational circuit in contrast to the exponential relation in other FHE frameworks. Moreover, the rounding operation is particularly suitable for machine learning applications, where the precision of the model parameters is of less concern. The second feature of this framework is that it supports the packing operation, where one ciphertext can encode a set of plaintext (parameters). As a result, operation on ciphertext is equivalent to doing parallel computation for the plaintext, which improves the training speed. 

We now briefly overview the recent development of HE-based data mining, which can be looked from multiple angles:
From the perspective of HE frameworks, partial HE frameworks \cite{paillier1999public} are widely used in early days for their simplicity, however, their applications are greatly restricted due to limited capabilities, \emph{e.g.}, applications based on the paillier framework \cite{paillier1999public} need to avoid encrypted multiplication by careful algorithmic design \cite{hardy2017private, cheng2019secureboost}. In the era of FHE, it is possible to train arbitrary model completely over the encrypted data in theory~\cite{carpov2019privacy, crawford2018doing}, however, the complexity becomes the major bottleneck and prevents the full exploitation of FHE's ability;
From the perspective of data mining models, most previous research focuses on shallow models, \emph{e.g.} logistic regression \cite{crawford2018doing, bonte2018privacy, carpov2019privacy, aono2016scalable, cheon2018ensemble}, decision tree \cite{cheng2019secureboost} and clustering \cite{cheon2019towards, jaschke2018unsupervised}. As for the recent popularized deep models, there is effort on making inference over the private data \cite{brutzkus2018low, gilad2016cryptonets, juvekar2018gazelle}, but not much on training. The major difficulty comes from the high complexity. On the one hand, the computation circuit of the gradient evaluation is much deeper compared to that of the shallow models, on the other hand, deep models typically need more iterations to converge. The two factors combined together make it impractical to train a deep model over the current HE system. 
Finally, there is also effort trying to make adaptions between the HE framework and the data mining models to make them more suitable for the other. In the HE side, \citet{jiang2018secure, mishra2018fast}~propose fast matrix-vector multiplication  kernels for the acceleration of HE-based data mining, and \citet{crawford2018doing} makes customized optimization for frequently used homomorphic operations by data mining like comparisons and inversion. In the model side, there are two main directions. One is to reduce the depth of computational circuit, \emph{e.g.} \citet{cheon2018ensemble} proposes to use ensemble methods so that each model can converge within less iterations and as a result, to reduce the circuit depth. The other is to linearize the machine learning models using methods like Taylor expansion or regression \cite{han2018efficient, bonte2018privacy} so as to alleviate the lack of efficient non-linear operations in most HE frameworks.

\section{Preliminary}
\label{sec:preliminary}
In this section, we briefly introduce some results that our framework is built upon. Firstly we introduce the Fully-HE framework HEAAN, which is composed of three main components, \emph{i.e.} encode/decode, encrypt/decrypt and basic homomorphic operations (addition, multiplication \emph{et. al.}). We briefly review them as follows (refer to \citet{cheon2017homomorphic} for more detailed description of each component):
\begin{itemize}
    \item \textbf{Encode/Decode} process transforms the plaintext, a vector in $R^N$, to/from a polynomial in the $N_{th}$ order cyclotomic polynomial ring (equivalently, the coefficients-vector of the polynomial) by the Canonical Embedding technique;
    \item \textbf{Encrypt/Decrypt} process transforms the plaintext to/from ciphertext which both are a cyclotomic polynomial. The public key and private key are needed in this process and are also cyclotomic polynomials. The security of this process is guaranteed by the hardness of Ring Learning-with-Errors (R-LWE) problem;
    \item \textbf{Homomorphic operations}
    supported by the framework are addition, multiplication, bootstrap and scaling operation. Bootstrap is for refreshing the noisy ciphertext and scaling performs homomorphic rounding operation so as to keep the modulus of the ciphertext linear with respect to the depth of the computational circuit.
\end{itemize}
We also mention two important complexity parameters of the framework. One is the order $N$ of the cyclotomic polynomial ring, which decides the number of plaintext we can pack into one ciphertext. The other is the magnitude $M$ of the ciphertext modulus which controls the complexity of basic homomorphic operations. Roughly speaking, we want to maximize $N$ and minimize $M$ to accelerate the computation, but $M$ and $N$ are positively correlated, so there is a trade-off between the two.

In \cite{han2018efficient}, the authors train a logistic regression model based on HEAAN. In this paper, \cite{han2018efficient} is referred as the centralized HE learner and will be used as the baseline for comparison. We briefly mention two useful tricks proposed by \citet{han2018efficient}. Firstly, the authors design a customized data packing schema for logistic regression. More specifically, suppose we have a $N$ by $D$ data matrix $X$, where $N$ is the number of samples and $D$ denotes the sample dimension. Then to encrypt $X$, we divide it into $l\times h$ small blocks and encrypt each block with one ciphertext. The parameter vector is grouped and encrypted using the similar way. As mentioned in \cite{han2018efficient}, the proposed packing method outperforms other more straightforward ways, \emph{e.g.} packing by rows or columns, in terms of the training speed. Next, the authors propose to use regression to linearize the nonlinear functions such as the Sigmoid function which contrasts with the usual Taylor expansion approach. The authors show the regression outperforms the Taylor expansion way. The regression way works as follows: suppose we want to fit a nonlinear function $f(x)$ with $M_{th}$ order polynomial $p(x) = \sum_{i=1}^{M}\alpha_i x^i$, then we solve the following high-order least square problem:
\begin{equation}
    \underset{\alpha \in R^{M}}{min} \sum_{j=1}^{N}\frac{1}{2}(f(x_j) - p(x_j))^2
\end{equation}
where $\{x_j\}_{j\in[N]}$ are randomly sampled. In this paper, we follow this idea to approximate other nonlinear activation functions such as the hinge loss function used in the SVM algorithm.

\section{Framework Design and Algorithm Implementation}
\label{sec:framework-design}
\begin{figure*}[ht]
  \centering
  \includegraphics[trim=40 0 10 0, clip, width=1.4\columnwidth]{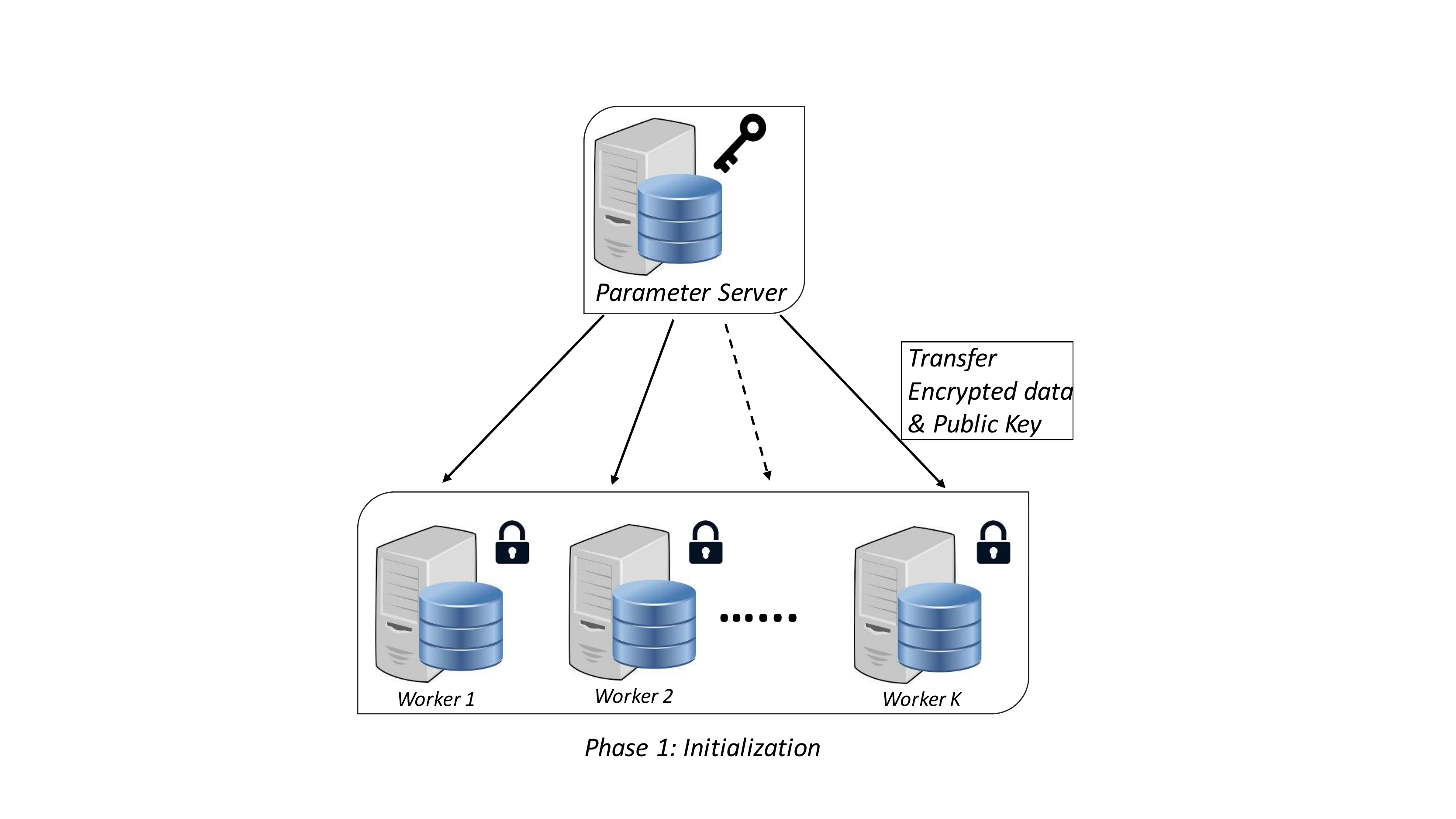}
 \hspace{-1in} 
 \includegraphics[trim=190 0 200 0, clip, width=0.85\columnwidth]{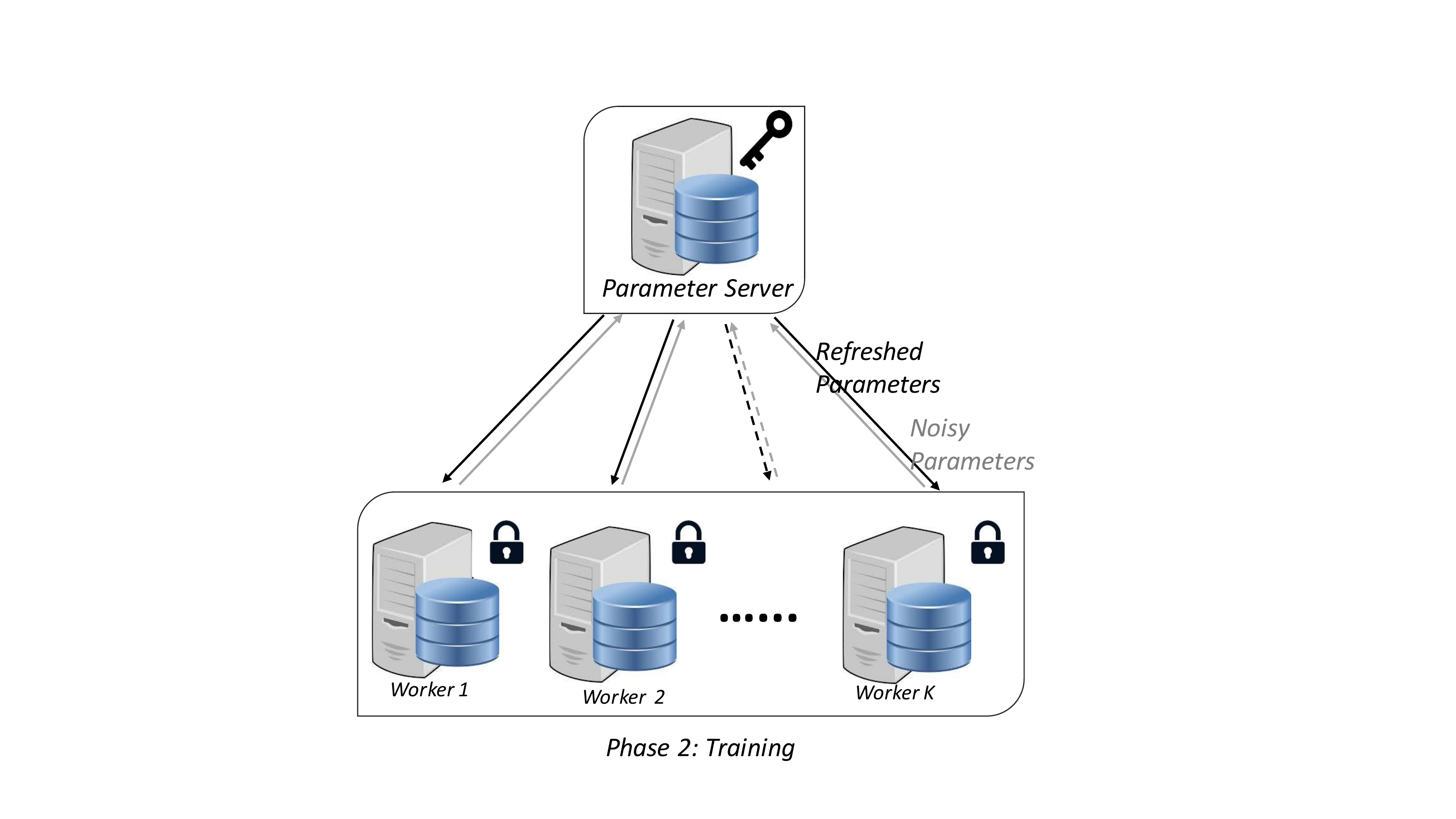}
  \caption{Two phases of our Training Framework. Left Figure shows initialization phase,
  and right figure shows the training phase.
}
\label{fig:framework}
\end{figure*}

In this section, we introduce our framework which enables faster secure data mining via distributed homomorphic encryption. The structure of our framework is shown in Figure~\ref{fig:framework}. We use the worker and parameter-server model for the distributed training. In our framework, we assume the parameter server is honest and hold the private key for encryption and decryption, while the workers perform the training jobs in a privacy preserving fashion, \emph{i.e.} the workers just have access to the encrypted data and the encrypted model parameters. More precisely, the framework includes two phases: the initialization phase and the training phase as shown in Figure~\ref{fig:framework}. In the initialization phase, the parameter server first generates the public and secret key pair and encrypts the data using the generated keys. Then it partitions the data and distributes the data to the corresponding worker. In the training phase, each worker performs training over the encrypted data until that the noise of the encrypted parameter reaches some predefined threshold, then the worker will send the encrypted parameter back to the parameter server and request update. In the parameter server side, it waits for the request of the worker. When it receives an update request, the parameter server decrypts the parameter and makes update to its own copy of the parameter, then it will send the encrypted version of the new parameter back to the worker.

Note that our framework is agnostic with specific HE frameworks, data mining models or distributed learning frameworks.
Our framework can boost the training speed tremendously as long as the bootstrap operation is the bottleneck of the application. 
But for the sake of demonstration, we consider the case of learning a linear model based on the HEAAN framework in a synchronous fashion.
More specifically, we study the following linear model:
\begin{equation}
\label{eq:obj}
\begin{split}
    &\underset{\omega\in\Omega\subset R^d}{min} f(\omega) \\
    =&\underset{\omega\in\Omega\subset R^d}{min} \sum_{i=1}^{n}L(y_i\times f(x_i)) + R(\omega)
\end{split}
\end{equation}
where $\{x_i,y_i\}_{i\in[n]}$ denotes the training data-set, $f(x_i) = \omega x_i$, and $\omega$ is the parameter. $L$ denotes the loss function, typical choices are summarize in Table~\ref{tab:loss}. $R$ is some regularization function such as $L_p$~norm regularization. Now we describe a method to train the above models based on our framework.

\begin{figure}[ht]
  \centering
  \includegraphics[width=0.9\columnwidth]{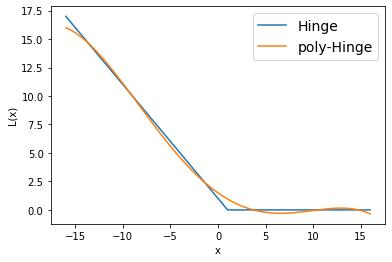}
  \caption{Hinge loss and its polynomial approximation based on linear regression. The blue curve is the hinge loss and the orange curve is its approximation.}
  \label{fig:appro}
\end{figure}

Suppose we use the gradient-based method to optimize Eq.~(\ref{eq:obj}), the update rule for $\omega$ is as follows:
\begin{equation}
\label{eq:update}
    \omega_{k+1} = \omega_{k} -\eta \nabla f(\omega_k)
\end{equation}
where $\eta$ is the learning rate and $\nabla f(\omega_k)$ denotes the gradient of $f$ with respect to $\omega_k$. We can derive the explicit form of $\nabla f(\omega_k)$ for linear models:
\begin{equation}
    \nabla f(\omega) = \sum_{i=1}^{n}y_i \nabla L \times x_i + \nabla R
\end{equation}
Where $\nabla L$ and $\nabla R$ denote the gradients of L and R respectively. Since the core step of learning a linear model is Eq.~(\ref{eq:update}), we need to linearize it for training under homomorphic encryption. By simple observation, the non-linearity mainly comes from $\nabla L$, we follow the method proposed by \citet{han2018efficient} to linearize it. We derive the polynomial approximation of the losses of Table~\ref{tab:loss} in Table~\ref{tab:lin}. As a example, we also plot the hinge loss and its polynomial approximation based on regression in Figure~\ref{fig:appro}.

\begin{table}
\setlength{\tabcolsep}{10pt}
  \caption{Loss Functions for Different Linear Models \cite{friedman2001elements}}
  \label{tab:loss}
  \begin{tabular}{ccl}
    \toprule
    Loss Function&$L(y,f(x))$\\
    \midrule
    Binomial Deviance & $\log(1+\exp(-yf(x)))$\\
    SVM Hinge Loss & $[1 - yf(x)]_{+}$\\
    Huber Loss &\begin{tabular}{@{}c@{}}$-4yf(x),\ yf(x) < -1$; \\ $[1 - yf(x)]_{+}^2$\ otherwise\end{tabular} \\
  \bottomrule
\end{tabular}
\end{table}

\begin{table}
  \caption{Polynomial Approximation for the gradient of the Losses in Table~\ref{tab:loss}. 4-order polynomial approximation is used so as to get a 3-order approximation of the gradient of the losses.}
  \label{tab:lin}
  \begin{tabular}{ccl}
    \toprule
    Loss Function&$\hat{\nabla}{L}(x)$\\
    \midrule
    Binomial Deviance & $0.5-0.0843x+0.0002x^3 $\\
    SVM Hinge Loss & $0.5875-0.1005x + 0.0008x^2 -0.00039x^3$\\
    Huber Loss & $2-0.1311x + 0.00005x^3$ \\
  \bottomrule
\end{tabular}
\end{table}

With the linearization, the update step in Eq.~(\ref{eq:update}) can be reformulated as:
\begin{equation}
    \omega_{k+1} = \omega_{k} -\eta \hat{\nabla} f(\omega_k)
\end{equation}
where:
\begin{equation}
\label{eq:appr}
    \hat{\nabla} f(\omega) = \sum_{i}^{n}y_i \hat{\nabla} L \times x_i + \nabla R
\end{equation}
and $\hat{\nabla} L$ is denoted in Table~\ref{tab:lin}, what's more, we assume the regularization function is smooth (polynomial) for simplicity here, but we can apply the same procedure to linearize $\nabla R$ as $\nabla L$ if necessary.

Now the update step only consists of addition and multiplication, we can define a computational circuit which can be evaluated by the HEAAN framework. The base learning algorithm used by every worker is shown in Algorithm~\ref{alg:para-server}. A key innovation here is that we avoid the use of bootstrap operation, instead we send the encrypted parameter $\pi(\omega)$ back to the parameter server for refreshing whenever the noise encoded in it goes beyond a threshold. Empirically, we simply refresh $\pi(\omega)$ every $l$ iterations as shown in Line 8 - 10 of Algorithm~\ref{alg:para-server}. Note this is critical in improving the performance of the framework and we will show later in the experimental results section that this leads to more than 10$\times$ training speed boost compared to the central version in \citet{han2018efficient}. What's more, we pack multiple training samples into one ciphertext using the pack technique in \cite{han2018efficient} to make use of the parallel capability of the HEAAN framework. As for the parameter server, it performs the function of combining the results from each worker and refreshing corrupted parameters on the request of the worker, its procedure is summarized in Algorithm~\ref{alg:central-server}.

\begin{algorithm}[tb]
   \caption{Training Algorithm for Worker}
   \label{alg:para-server}
\begin{algorithmic}[1]
   \STATE {\bfseries Input:} Learning rate\ $\eta$, public key\ $pk$, number of iterations $l$ to update the local weights, number of training iterations $K$; 
   \STATE {\bfseries Output:} Converged encrypted paramter $\pi(\omega_K)$;
   \STATE {\bfseries Initialization Phase: }
   \STATE Receive data from the parameter server, \emph{i.e.} encrypted data-set $\{\pi(x_i),\pi(y_i)\}_{i\in[N]}$, public key\ $pk$ and initial encrypted parameter $\pi(\omega_0)$;
   \STATE {\bfseries Training Phase: }
       \FOR{k = 1 to K}
       \STATE update $\pi(\omega)$ by evaluating Eq.~(\ref{eq:appr}) homomorphically with $pk$;
       \IF{k \% $l$ = 0}
       \STATE send $\pi(\omega)$ to the parameter server and wait until the parameter server sends back the updated weights;
       \ENDIF
        \ENDFOR
\end{algorithmic}
\end{algorithm}

\begin{algorithm}[tb]
   \caption{Training Algorithm for Parameter Server}
   \label{alg:central-server}
\begin{algorithmic}[1]
   \STATE {\bfseries Input:} Training set $\{x_i,y_i\}_{i\in[N]}$, initial parameter $\omega_0$; 
   \STATE {\bfseries Output:} Converged parameter $\omega_K$;
   \STATE {\bfseries Initialization Phase: }
   \STATE Generate public and secret keys, \emph{i.e.} $pk$ and $sk$, encrypt training set $\{x_i,y_i\}_{i\in[N]}$ and initial parameter $\omega_0$ with $sk$;
   \STATE Send the encrypted data-set $\{\pi(x_i),\pi(y_i)\}_{i\in[N]}$, public key\ $pk$ and initial encrypted parameter $\pi(\omega_0)$ to each worker;
   \STATE {\bfseries Training Phase: }
       \WHILE{the workers are running}
       \IF{receive request from workers}
       \STATE Decrypt the encrypted parameter $\pi(\omega)_{param}$ received from the worker to get $\omega_{param}$;
       \STATE Update its own $\omega$ with $\omega_{param}$;
       \STATE Encrypt $\omega$ and send back to the worker;
       \ENDIF
       \ENDWHILE
\end{algorithmic}
\end{algorithm}

\section{Experimental Results}
\label{sec:experimental-results}
In this section, we present the experimental results. To show the effectiveness of our framework, we compare with the centralized HE learning framework \cite{han2018efficient}, and make ablation studies of the key factors that influence the performance of our framework. In the figure, we use \lq Plain\rq\ to represent results without encryption; use \lq Centralized\rq\ to show the experimental results based on the centralized learning framework.

More specifically, in section~\ref{distributed-learning}, we test under the distributed learning setting, where we train different linear models on a simplified MNIST data-set and a Financial Data-set \cite{han2018efficient}, and also do various ablation studies to study the influence of different factors (hyper-parameters) to the performance of our framework. In section~\ref{federated-learning}, we test on a federated learning data-set called FEMNIST~\cite{caldas2018leaf} to show that our framework is also effective under the federated learning scenario.

The experiments are conducted over an Intel Xeon E5-2683 v4~@~2.10GHz server with 64 cores. We use OpenMPI \cite{graham2005open} library to simulate the communication between workers and the parameter server. OpenMPI is an open source Message Passing Interface(MPI) library widely used in high performance computing. What's more, we use a C++ implementation of HEAAN~\footnote{https://github.com/snucrypto/HEAAN} to perform HE related computations, such as homomorphic addition and multiplication.

\subsection{Distributed Learning}
\label{distributed-learning}
In this subsection, we test our framework under distributed learning settings. More specifically, we learn different linear models over a simplified MNIST data-set and a Financial data-set. The data-set construction process is as follows: we extract the digit 3 and 8 from MNIST data-set to construct a data-set with 11982 training samples and 1984 validation samples. Then we randomly split the training set to construct the training set and the test set (the test set is kept by the parameter server for evaluation). We also sub-sample the digit images to 14$\times$14. As for the financial data-set, we make use of the public subset of the original data-set mentioned in \cite{han2018efficient} which includes the credit information of 30000 individuals and 24 features for each individual. We split the data-set to a training set of 27000 samples and a validation set of 3000 samples, and then randomly split the training set further to construct data-set for each worker. 

\begin{figure}[ht]
  \centering
  \includegraphics[width=0.9\columnwidth]{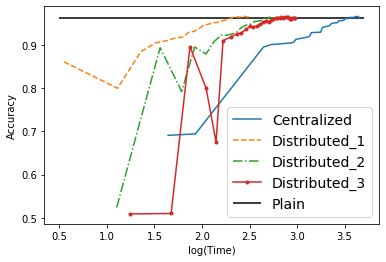}
  \caption{Validation accuracy with respect to the training time (log scale) of a logistic regression model trained on the simplified MNIST data-set. \lq Distributed-k\rq\ means workers refresh the parameters every k iterations for our framework.}
  \label{fig:time}
\end{figure}

Now we present some experimental results. In Figure~\ref{fig:time}, we plot results of the validation accuracy with respect to the training time (log scale) when we train a logistic regression model over the MNIST data-set. 
As shown in the figure, both our distributed learning algorithm and the centralized learning algorithm reach the accuracy of around 96.5\% which is the same as the plain accuracy, however, when comparing the training time, our framework converges much faster than the centralized version. The centralized framework consumes around 4000-5000s to converge, while our framework typically needs around 300-400s. We have carefully tuned the centralized framework to reach the best time performance, more specifically, we perform bootstrap every 3 iterations for the centralized framework to reach the best comprise between the magnitude of the ciphertext and the frequency of bootstrap.
As for our framework, we varies the frequency that workers refresh their parameters. Apparently, the more frequent the worker refreshes its parameter, the higher communication overhead it leads to, however, the variant that refreshes the parameter every iteration surpasses all the other variants, although with the largest communication overhead. This actually shows that communication cost is negligible even compared to the cost of basic homomorphic operations, let alone the more expensive bootstrap operation.

\begin{figure}[ht]
  \centering
  \includegraphics[width=0.9\columnwidth]{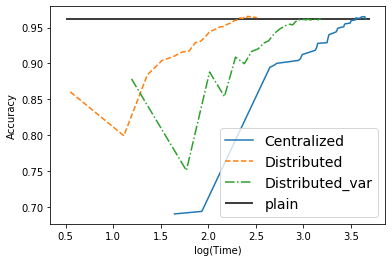}
  \caption{Validation accuracy with respect to training time (log scale) of a logistic regression model trained on the simplified MNIST data-set. \lq Distributed\rq\ is the results of our framework, while \lq Distributed-var\rq\ means the variant of our framework where the magnitude of ciphertext modulus is kept the same.}
  \label{fig:modulus}
\end{figure}

We now analyze our framework in more details. The acceleration actually comes from two parts. Firstly, our framework saves a great deal of the computational time by not performing the super expensive bootstrap operation. 
Furthermore, an indirect effect of eliminating the bootstrap operation is the smaller magnitude of the ciphertext modulus. This is because the magnitude of the ciphertext modulus is positively correlated with the depth of the computational circuit, while bootstrap operation takes up a significant part of the circuit depth. Naturally, we can use smaller ciphertext modulus when the computational circuit becomes shallower. Actually, the computational time for other basic homomorphic operations also decreases when we are able to use a smaller ciphertext modulus. These all benefit the training speed of our framework. To illustrate more about the effect of the ciphertext modulus, we train a variant where we only eliminate bootstrap operation but keep the magnitude of the ciphertext modulus the same. The results are shown in Figure~\ref{fig:modulus}. As we can see, simply eliminating the bootstrap operation can save around half of the computational time, which is still at the order of $10^3$s, however, if we also make use of the saved ciphertext modulus, the training time can be further decreased to the order $10^2$s. To sum up, our framework avoids the expensive bootstrap operation and also decrease the complexity of other basic operations by reducing the ciphertext modulus. Though with the cost of communication, this is negligible as shown by the experimental results.

\begin{figure}[ht]
  \centering
  \includegraphics[width=0.9\columnwidth]{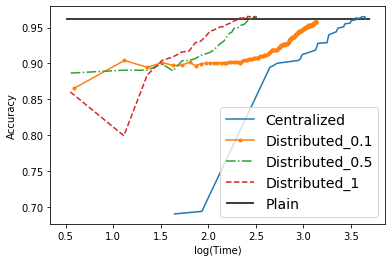}
  \caption{Validation accuracy with respect to the training time (log scale) of a logistic regression model trained on MNIST data-set. \lq Distributed-k\rq\ means the learning rate used by our framework is k. }
  \label{fig:lr}
\end{figure}

\begin{figure}[ht]
  \centering
  \includegraphics[width=0.9\columnwidth]{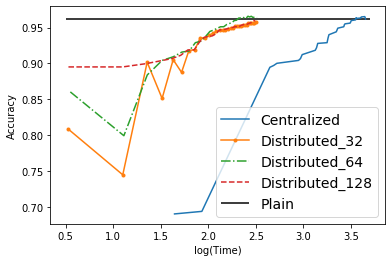}
  \caption{Validation accuracy with respect to the training time (log scale) of a logistic regression model trained on MNIST data-set. \lq Distributed-k\rq\ means that the batch size used by our framework is k.}
  \label{fig:batchsize}
\end{figure}

\begin{figure}[ht]
  \centering
  \includegraphics[width=0.9\columnwidth]{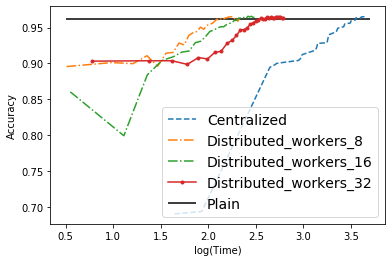}
  \caption{Validation accuracy with respect to training time (log scale) of a logistic regression model trained on MNIST data-set. \lq Distributed-worker-k\rq\ represents the number of workers used is k.}
  \label{fig:worker}
\end{figure}

To illustrate the robustness of our framework, we train different linear models on the MNIST data-set with the polynomial approximation computed in Table~\ref{tab:lin}. The best validation accuracy and the corresponding training time for each model is shown in Table \ref{tab:acc-time}. The accuracy and training time is roughly the same for each loss, which shows the robustness of our framework. 
Next, we vary the hyper-parameters to test their effect to the learning process. The results are shown in Figure~\ref{fig:lr} and Figure~\ref{fig:batchsize}. The hyper-parameters of the experiments related to the centralized framework is kept the same as previous experiments. For the experiments in Figure~\ref{fig:lr}, we fix batch-size at 64 and vary the learning rate. While for the experiments in Figure~\ref{fig:batchsize}, we fix the learning rate at 1 and change the batch-size. As shown by the experimental results, learning rate 1 and batch-size 128 is a reasonable choice with both fast and stable convergence. What's more, we also compare the influence of the number of workers, the results are shown in Figure~\ref{fig:worker}. We can see that more workers lead to higher communication overhead and slightly longer training time.

\begin{table}
\setlength{\tabcolsep}{12pt}
  \caption{Validation Accuracy and Training Time for Different Linear Models trained on MNIST data-set}
  \label{tab:acc-time}
  \begin{tabular}{ccl}
    \toprule
    Loss Function&Accuracy&Time(s)\\
    \midrule
    Binomial Deviance & 96.52 & 323.44\\
    SVM Hinge Loss & 96.42 & 333.74\\
    Huber & 96.57 & 333.65\\
  \bottomrule
\end{tabular}
\end{table}

We also test our framework on a bigger and more realistic data-set, \emph{i.e.} the financial data-set as mentioned before. We compare the validation accuracy and training time of our framework with that of the centralized framework and show the results in the Table~\ref{tab:fin}. Note that for this bigger and more complicated data-set, our model can achieve roughly the same accuracy (even slightly better) with the centralized model while still remaining running 10$\times$ times faster.

\begin{table}
\setlength{\tabcolsep}{12pt}
  \caption{Validation Accuracy and Training Time of Logistic Regression Model over Financial Data-set}
  \label{tab:fin}
  \begin{tabular}{ccl}
    \toprule
    Framework&Accuracy&Time(s)\\
    \midrule
    Distributed & 79.62&303.8\\
    Centralized & 79.42&4059.68\\
    Plain & 79.42&-\\
  \bottomrule
\end{tabular}
\end{table}

\subsection{Federated Learning}
\label{federated-learning}
\begin{figure}[ht]
  \centering
  \includegraphics[width=.4\columnwidth]{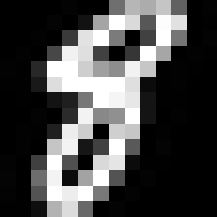}
  \qquad
  \includegraphics[width=.4\columnwidth]{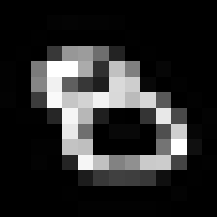}
  \caption{Two samples of digits from FEMNIST dataset.}
  \label{fig:sample}
\end{figure}

In this subsection, we test our framework under the federated learning setting. The concept of Federated learning \cite{konevcny2016federated1, konevcny2016federated2, yang2019federated} is proposed for learning over data-sets that are distributed across multiple devices and the distribution of which may be heterogeneous. Security and privacy of the user data is a great challenge in Federated Learning, which is a great match of our fast HE-based data mining framework. In this paper, we simulate the federated learning scenario with the FEMNIST data-set \cite{caldas2018leaf}. FEMNIST is generated from the MNIST data-set by partitioning the data-set according to the writer of each character. Since each writer has a special writing style, the underlying distribution of the characters written by them are different and therefore, FEMNIST is a good choice to simulate the federated learning situation. More specifically, we construct the data-set as follows: we randomly select 50 writers for each worker and construct a data-set with the characters they write. In Figure~\ref{fig:sample}, we show two samples of the digits written by two different writers. The digit 8 in the left figure is slimmer and orients to the right, while the digit in the right is fatter and orients to the left. As for the data-set used by the centralized learning framework, we combine all of the users' characters together to form a data-set.

\begin{figure}[ht]
  \centering
  \includegraphics[width=0.9\columnwidth]{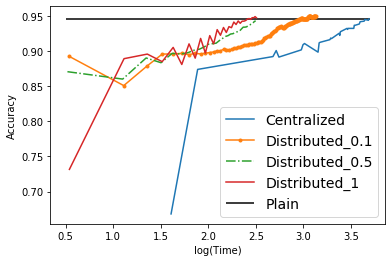}
  \caption{Validation accuracy with respect to training time (log scale) of a logistic regression model trained on FEMNIST data-set. \lq Distributed-k\rq\ means the workers use learning rate k. We use batch-size 64 in this experiment.}
  \label{fig:fed-lr}
\end{figure}

\begin{figure}[ht]
  \centering
  \includegraphics[width=0.9\columnwidth]{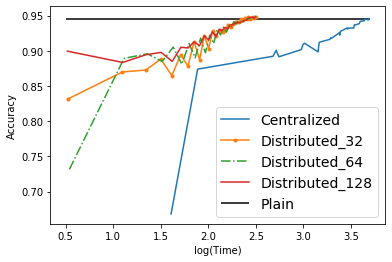}
  \caption{Validation accuracy with respect to training time (log scale) of a logistic regression model trained on FEMNIST data-set. \lq Distributed-k\rq\ means the workers use batch-size k, we use learning rate 1 in this experiment.}
  \label{fig:fed-batchsize}
\end{figure}

As in the distributed learning case, we learn different models over FEMNIST data-set and the accuracy and training time is shown in Table~\ref{tab:fed-acc-time}. Their performance is roughly the same and the SVM model gets a slightly worse accuracy. What's more, we vary the learning rate and batch-size to show the robustness of our framework. The results are shown in Figure~\ref{fig:fed-lr} and Figure~\ref{fig:fed-batchsize}. The results show that our framework achieves the same training speed boost \emph{i.e.} 10$\times$ times under federated learning settings.

\begin{table}
\setlength{\tabcolsep}{12pt}
  \caption{Validation Accuracy and Training Time for Different Linear Models trained on FEMNIST data-set}
  \label{tab:fed-acc-time}
  \begin{tabular}{ccl}
    \toprule
    Loss Function&Accuracy&Time(s)\\
    \midrule
    Binomial Deviance & 94.64&327.09\\
    SVM Hinge Loss & 93.66&337.63\\
    Huber Loss &94.91&327.01 \\
  \bottomrule
\end{tabular}
\end{table}

\section{Conclusion}
\label{sec:concluson}
In this paper, we propose a distributed learning framework to drastically improve the training speed of HE-based data mining algorithms. For example, we train a logistic regression model over a data-set with 10 thousand records within 5 minutes, while the centralized framework needs almost 2 hours. This performance boost is achieved by eliminating the bootstrap operation with the cost of communication between the workers and the parameter server. What's more, our framework is agnostic with different HE frameworks, distributed learning frameworks or specific models. We leave it as a future work to test our framework in larger scale, however, we note that the computational time of the basic arithmetic operation (especially the ciphertext multiplication) is still a bottleneck which is several orders slower than the plain operation. With the development of HE, the performance of our framework can be improved further. Finally, the model maintains privacy since it performs evaluation in the encryption field, however, the model may still contain some private information if we decrypt it, one possible remedy is to combine our framework with other privacy preserving techniques like differential privacy.

\bibliographystyle{ACM-Reference-Format}
\bibliography{sample-sigconf}


\begin{thebibliography}{37}


\ifx \showCODEN    \undefined \def \showCODEN     #1{\unskip}     \fi
\ifx \showDOI      \undefined \def \showDOI       #1{#1}\fi
\ifx \showISBNx    \undefined \def \showISBNx     #1{\unskip}     \fi
\ifx \showISBNxiii \undefined \def \showISBNxiii  #1{\unskip}     \fi
\ifx \showISSN     \undefined \def \showISSN      #1{\unskip}     \fi
\ifx \showLCCN     \undefined \def \showLCCN      #1{\unskip}     \fi
\ifx \shownote     \undefined \def \shownote      #1{#1}          \fi
\ifx \showarticletitle \undefined \def \showarticletitle #1{#1}   \fi
\ifx \showURL      \undefined \def \showURL       {\relax}        \fi
\providecommand\bibfield[2]{#2}
\providecommand\bibinfo[2]{#2}
\providecommand\natexlab[1]{#1}
\providecommand\showeprint[2][]{arXiv:#2}

\bibitem[\protect\citeauthoryear{Aono, Hayashi, Trieu~Phong, and Wang}{Aono
  et~al\mbox{.}}{2016}]%
        {aono2016scalable}
\bibfield{author}{\bibinfo{person}{Yoshinori Aono}, \bibinfo{person}{Takuya
  Hayashi}, \bibinfo{person}{Le Trieu~Phong}, {and} \bibinfo{person}{Lihua
  Wang}.} \bibinfo{year}{2016}\natexlab{}.
\newblock \showarticletitle{Scalable and secure logistic regression via
  homomorphic encryption}. In \bibinfo{booktitle}{\emph{Proceedings of the
  Sixth ACM Conference on Data and Application Security and Privacy}}.
  \bibinfo{pages}{142--144}.
\newblock


\bibitem[\protect\citeauthoryear{Bonte and Vercauteren}{Bonte and
  Vercauteren}{2018}]%
        {bonte2018privacy}
\bibfield{author}{\bibinfo{person}{Charlotte Bonte} {and}
  \bibinfo{person}{Frederik Vercauteren}.} \bibinfo{year}{2018}\natexlab{}.
\newblock \showarticletitle{Privacy-preserving logistic regression training}.
\newblock \bibinfo{journal}{\emph{BMC medical genomics}} \bibinfo{volume}{11},
  \bibinfo{number}{4} (\bibinfo{year}{2018}), \bibinfo{pages}{86}.
\newblock


\bibitem[\protect\citeauthoryear{Brakerski, Gentry, and
  Vaikuntanathan}{Brakerski et~al\mbox{.}}{2014}]%
        {brakerski2014leveled}
\bibfield{author}{\bibinfo{person}{Zvika Brakerski}, \bibinfo{person}{Craig
  Gentry}, {and} \bibinfo{person}{Vinod Vaikuntanathan}.}
  \bibinfo{year}{2014}\natexlab{}.
\newblock \showarticletitle{(Leveled) fully homomorphic encryption without
  bootstrapping}.
\newblock \bibinfo{journal}{\emph{ACM Transactions on Computation Theory
  (TOCT)}} \bibinfo{volume}{6}, \bibinfo{number}{3} (\bibinfo{year}{2014}),
  \bibinfo{pages}{1--36}.
\newblock


\bibitem[\protect\citeauthoryear{Brakerski and Vaikuntanathan}{Brakerski and
  Vaikuntanathan}{2011}]%
        {brakerski2011fully}
\bibfield{author}{\bibinfo{person}{Zvika Brakerski} {and}
  \bibinfo{person}{Vinod Vaikuntanathan}.} \bibinfo{year}{2011}\natexlab{}.
\newblock \showarticletitle{Fully homomorphic encryption from ring-LWE and
  security for key dependent messages}. In \bibinfo{booktitle}{\emph{Annual
  cryptology conference}}. Springer, \bibinfo{pages}{505--524}.
\newblock


\bibitem[\protect\citeauthoryear{Brakerski and Vaikuntanathan}{Brakerski and
  Vaikuntanathan}{2014}]%
        {brakerski2014efficient}
\bibfield{author}{\bibinfo{person}{Zvika Brakerski} {and}
  \bibinfo{person}{Vinod Vaikuntanathan}.} \bibinfo{year}{2014}\natexlab{}.
\newblock \showarticletitle{Efficient fully homomorphic encryption from
  (standard) LWE}.
\newblock \bibinfo{journal}{\emph{SIAM J. Comput.}} \bibinfo{volume}{43},
  \bibinfo{number}{2} (\bibinfo{year}{2014}), \bibinfo{pages}{831--871}.
\newblock


\bibitem[\protect\citeauthoryear{Brutzkus, Elisha, and Gilad-Bachrach}{Brutzkus
  et~al\mbox{.}}{2018}]%
        {brutzkus2018low}
\bibfield{author}{\bibinfo{person}{Alon Brutzkus}, \bibinfo{person}{Oren
  Elisha}, {and} \bibinfo{person}{Ran Gilad-Bachrach}.}
  \bibinfo{year}{2018}\natexlab{}.
\newblock \showarticletitle{Low latency privacy preserving inference}.
\newblock \bibinfo{journal}{\emph{arXiv preprint arXiv:1812.10659}}
  (\bibinfo{year}{2018}).
\newblock


\bibitem[\protect\citeauthoryear{Caldas, Wu, Li, Kone{\v{c}}n{\`y}, McMahan,
  Smith, and Talwalkar}{Caldas et~al\mbox{.}}{2018}]%
        {caldas2018leaf}
\bibfield{author}{\bibinfo{person}{Sebastian Caldas}, \bibinfo{person}{Peter
  Wu}, \bibinfo{person}{Tian Li}, \bibinfo{person}{Jakub Kone{\v{c}}n{\`y}},
  \bibinfo{person}{H~Brendan McMahan}, \bibinfo{person}{Virginia Smith}, {and}
  \bibinfo{person}{Ameet Talwalkar}.} \bibinfo{year}{2018}\natexlab{}.
\newblock \showarticletitle{Leaf: A benchmark for federated settings}.
\newblock \bibinfo{journal}{\emph{arXiv preprint arXiv:1812.01097}}
  (\bibinfo{year}{2018}).
\newblock


\bibitem[\protect\citeauthoryear{Carpov, Gama, Georgieva, and
  Troncoso-Pastoriza}{Carpov et~al\mbox{.}}{2019}]%
        {carpov2019privacy}
\bibfield{author}{\bibinfo{person}{Sergiu Carpov}, \bibinfo{person}{Nicolas
  Gama}, \bibinfo{person}{Mariya Georgieva}, {and}
  \bibinfo{person}{Juan~Ram{\'o}n Troncoso-Pastoriza}.}
  \bibinfo{year}{2019}\natexlab{}.
\newblock \showarticletitle{Privacy-preserving semi-parallel logistic
  regression training with Fully Homomorphic Encryption.}
\newblock \bibinfo{journal}{\emph{IACR Cryptology ePrint Archive}}
  \bibinfo{volume}{2019} (\bibinfo{year}{2019}), \bibinfo{pages}{101}.
\newblock


\bibitem[\protect\citeauthoryear{Challa}{Challa}{2020}]%
        {challa2020homomorphic}
\bibfield{author}{\bibinfo{person}{Ratnakumari Challa}.}
  \bibinfo{year}{2020}\natexlab{}.
\newblock \showarticletitle{Homomorphic Encryption: Review and Applications}.
\newblock In \bibinfo{booktitle}{\emph{Advances in Data Science and
  Management}}. \bibinfo{publisher}{Springer}, \bibinfo{pages}{273--281}.
\newblock


\bibitem[\protect\citeauthoryear{Champagne and Lee}{Champagne and Lee}{2010}]%
        {champagne2010scalable}
\bibfield{author}{\bibinfo{person}{David Champagne} {and}
  \bibinfo{person}{Ruby~B Lee}.} \bibinfo{year}{2010}\natexlab{}.
\newblock \showarticletitle{Scalable architectural support for trusted
  software}. In \bibinfo{booktitle}{\emph{HPCA-16 2010 The Sixteenth
  International Symposium on High-Performance Computer Architecture}}. IEEE,
  \bibinfo{pages}{1--12}.
\newblock


\bibitem[\protect\citeauthoryear{Cheng, Fan, Jin, Liu, Chen, and Yang}{Cheng
  et~al\mbox{.}}{2019}]%
        {cheng2019secureboost}
\bibfield{author}{\bibinfo{person}{Kewei Cheng}, \bibinfo{person}{Tao Fan},
  \bibinfo{person}{Yilun Jin}, \bibinfo{person}{Yang Liu},
  \bibinfo{person}{Tianjian Chen}, {and} \bibinfo{person}{Qiang Yang}.}
  \bibinfo{year}{2019}\natexlab{}.
\newblock \showarticletitle{Secureboost: A lossless federated learning
  framework}.
\newblock \bibinfo{journal}{\emph{arXiv preprint arXiv:1901.08755}}
  (\bibinfo{year}{2019}).
\newblock


\bibitem[\protect\citeauthoryear{Cheon, Kim, Kim, and Song}{Cheon
  et~al\mbox{.}}{2017}]%
        {cheon2017homomorphic}
\bibfield{author}{\bibinfo{person}{Jung~Hee Cheon}, \bibinfo{person}{Andrey
  Kim}, \bibinfo{person}{Miran Kim}, {and} \bibinfo{person}{Yongsoo Song}.}
  \bibinfo{year}{2017}\natexlab{}.
\newblock \showarticletitle{Homomorphic encryption for arithmetic of
  approximate numbers}. In \bibinfo{booktitle}{\emph{International Conference
  on the Theory and Application of Cryptology and Information Security}}.
  Springer, \bibinfo{pages}{409--437}.
\newblock


\bibitem[\protect\citeauthoryear{Cheon, Kim, Kim, and Song}{Cheon
  et~al\mbox{.}}{2018}]%
        {cheon2018ensemble}
\bibfield{author}{\bibinfo{person}{Jung~Hee Cheon}, \bibinfo{person}{Duhyeong
  Kim}, \bibinfo{person}{Yongdai Kim}, {and} \bibinfo{person}{Yongsoo Song}.}
  \bibinfo{year}{2018}\natexlab{}.
\newblock \showarticletitle{Ensemble method for privacy-preserving logistic
  regression based on homomorphic encryption}.
\newblock \bibinfo{journal}{\emph{IEEE Access}}  \bibinfo{volume}{6}
  (\bibinfo{year}{2018}), \bibinfo{pages}{46938--46948}.
\newblock


\bibitem[\protect\citeauthoryear{Cheon, Kim, and Park}{Cheon
  et~al\mbox{.}}{2019}]%
        {cheon2019towards}
\bibfield{author}{\bibinfo{person}{Jung~Hee Cheon}, \bibinfo{person}{Duhyeong
  Kim}, {and} \bibinfo{person}{Jai~Hyun Park}.}
  \bibinfo{year}{2019}\natexlab{}.
\newblock \showarticletitle{Towards a practical cluster analysis over encrypted
  data}. In \bibinfo{booktitle}{\emph{International Conference on Selected
  Areas in Cryptography}}. Springer, \bibinfo{pages}{227--249}.
\newblock


\bibitem[\protect\citeauthoryear{Crawford, Gentry, Halevi, Platt, and
  Shoup}{Crawford et~al\mbox{.}}{2018}]%
        {crawford2018doing}
\bibfield{author}{\bibinfo{person}{Jack~LH Crawford}, \bibinfo{person}{Craig
  Gentry}, \bibinfo{person}{Shai Halevi}, \bibinfo{person}{Daniel Platt}, {and}
  \bibinfo{person}{Victor Shoup}.} \bibinfo{year}{2018}\natexlab{}.
\newblock \showarticletitle{Doing real work with FHE: the case of logistic
  regression}. In \bibinfo{booktitle}{\emph{Proceedings of the 6th Workshop on
  Encrypted Computing \& Applied Homomorphic Cryptography}}.
  \bibinfo{pages}{1--12}.
\newblock


\bibitem[\protect\citeauthoryear{Dwork}{Dwork}{2008}]%
        {dwork2008differential}
\bibfield{author}{\bibinfo{person}{Cynthia Dwork}.}
  \bibinfo{year}{2008}\natexlab{}.
\newblock \showarticletitle{Differential privacy: A survey of results}. In
  \bibinfo{booktitle}{\emph{International conference on theory and applications
  of models of computation}}. Springer, \bibinfo{pages}{1--19}.
\newblock


\bibitem[\protect\citeauthoryear{Dwork, Roth, et~al\mbox{.}}{Dwork
  et~al\mbox{.}}{2014}]%
        {dwork2014algorithmic}
\bibfield{author}{\bibinfo{person}{Cynthia Dwork}, \bibinfo{person}{Aaron
  Roth}, {et~al\mbox{.}}} \bibinfo{year}{2014}\natexlab{}.
\newblock \showarticletitle{The algorithmic foundations of differential
  privacy}.
\newblock \bibinfo{journal}{\emph{Foundations and Trends{\textregistered} in
  Theoretical Computer Science}} \bibinfo{volume}{9}, \bibinfo{number}{3--4}
  (\bibinfo{year}{2014}), \bibinfo{pages}{211--407}.
\newblock


\bibitem[\protect\citeauthoryear{Friedman, Hastie, and Tibshirani}{Friedman
  et~al\mbox{.}}{2001}]%
        {friedman2001elements}
\bibfield{author}{\bibinfo{person}{Jerome Friedman}, \bibinfo{person}{Trevor
  Hastie}, {and} \bibinfo{person}{Robert Tibshirani}.}
  \bibinfo{year}{2001}\natexlab{}.
\newblock \bibinfo{booktitle}{\emph{The elements of statistical learning}}.
  Vol.~\bibinfo{volume}{1}.
\newblock \bibinfo{publisher}{Springer series in statistics New York}.
\newblock


\bibitem[\protect\citeauthoryear{Gentry}{Gentry}{2009}]%
        {gentry2009fully}
\bibfield{author}{\bibinfo{person}{Craig Gentry}.}
  \bibinfo{year}{2009}\natexlab{}.
\newblock \bibinfo{booktitle}{\emph{A fully homomorphic encryption scheme}}.
  Vol.~\bibinfo{volume}{20}.
\newblock \bibinfo{publisher}{Stanford university Stanford}.
\newblock


\bibitem[\protect\citeauthoryear{Gilad-Bachrach, Dowlin, Laine, Lauter,
  Naehrig, and Wernsing}{Gilad-Bachrach et~al\mbox{.}}{2016}]%
        {gilad2016cryptonets}
\bibfield{author}{\bibinfo{person}{Ran Gilad-Bachrach}, \bibinfo{person}{Nathan
  Dowlin}, \bibinfo{person}{Kim Laine}, \bibinfo{person}{Kristin Lauter},
  \bibinfo{person}{Michael Naehrig}, {and} \bibinfo{person}{John Wernsing}.}
  \bibinfo{year}{2016}\natexlab{}.
\newblock \showarticletitle{Cryptonets: Applying neural networks to encrypted
  data with high throughput and accuracy}. In
  \bibinfo{booktitle}{\emph{International Conference on Machine Learning}}.
  \bibinfo{pages}{201--210}.
\newblock


\bibitem[\protect\citeauthoryear{Graham, Woodall, and Squyres}{Graham
  et~al\mbox{.}}{2005}]%
        {graham2005open}
\bibfield{author}{\bibinfo{person}{Richard~L Graham},
  \bibinfo{person}{Timothy~S Woodall}, {and} \bibinfo{person}{Jeffrey~M
  Squyres}.} \bibinfo{year}{2005}\natexlab{}.
\newblock \showarticletitle{Open MPI: A flexible high performance MPI}. In
  \bibinfo{booktitle}{\emph{International Conference on Parallel Processing and
  Applied Mathematics}}. Springer, \bibinfo{pages}{228--239}.
\newblock


\bibitem[\protect\citeauthoryear{Han, Hong, Cheon, and Park}{Han
  et~al\mbox{.}}{2018}]%
        {han2018efficient}
\bibfield{author}{\bibinfo{person}{Kyoohyung Han}, \bibinfo{person}{Seungwan
  Hong}, \bibinfo{person}{Jung~Hee Cheon}, {and} \bibinfo{person}{Daejun
  Park}.} \bibinfo{year}{2018}\natexlab{}.
\newblock \showarticletitle{Efficient Logistic Regression on Large Encrypted
  Data.}
\newblock \bibinfo{journal}{\emph{IACR Cryptology ePrint Archive}}
  \bibinfo{volume}{2018} (\bibinfo{year}{2018}), \bibinfo{pages}{662}.
\newblock


\bibitem[\protect\citeauthoryear{Hardy, Henecka, Ivey-Law, Nock, Patrini,
  Smith, and Thorne}{Hardy et~al\mbox{.}}{2017}]%
        {hardy2017private}
\bibfield{author}{\bibinfo{person}{Stephen Hardy}, \bibinfo{person}{Wilko
  Henecka}, \bibinfo{person}{Hamish Ivey-Law}, \bibinfo{person}{Richard Nock},
  \bibinfo{person}{Giorgio Patrini}, \bibinfo{person}{Guillaume Smith}, {and}
  \bibinfo{person}{Brian Thorne}.} \bibinfo{year}{2017}\natexlab{}.
\newblock \showarticletitle{Private federated learning on vertically
  partitioned data via entity resolution and additively homomorphic
  encryption}.
\newblock \bibinfo{journal}{\emph{arXiv preprint arXiv:1711.10677}}
  (\bibinfo{year}{2017}).
\newblock


\bibitem[\protect\citeauthoryear{J{\"a}schke and Armknecht}{J{\"a}schke and
  Armknecht}{2018}]%
        {jaschke2018unsupervised}
\bibfield{author}{\bibinfo{person}{Angela J{\"a}schke} {and}
  \bibinfo{person}{Frederik Armknecht}.} \bibinfo{year}{2018}\natexlab{}.
\newblock \showarticletitle{Unsupervised machine learning on encrypted data}.
  In \bibinfo{booktitle}{\emph{International Conference on Selected Areas in
  Cryptography}}. Springer, \bibinfo{pages}{453--478}.
\newblock


\bibitem[\protect\citeauthoryear{Jiang, Kim, Lauter, and Song}{Jiang
  et~al\mbox{.}}{2018}]%
        {jiang2018secure}
\bibfield{author}{\bibinfo{person}{Xiaoqian Jiang}, \bibinfo{person}{Miran
  Kim}, \bibinfo{person}{Kristin Lauter}, {and} \bibinfo{person}{Yongsoo
  Song}.} \bibinfo{year}{2018}\natexlab{}.
\newblock \showarticletitle{Secure outsourced matrix computation and
  application to neural networks}. In \bibinfo{booktitle}{\emph{Proceedings of
  the 2018 ACM SIGSAC Conference on Computer and Communications Security}}.
  \bibinfo{pages}{1209--1222}.
\newblock


\bibitem[\protect\citeauthoryear{Juvekar, Vaikuntanathan, and
  Chandrakasan}{Juvekar et~al\mbox{.}}{2018}]%
        {juvekar2018gazelle}
\bibfield{author}{\bibinfo{person}{Chiraag Juvekar}, \bibinfo{person}{Vinod
  Vaikuntanathan}, {and} \bibinfo{person}{Anantha Chandrakasan}.}
  \bibinfo{year}{2018}\natexlab{}.
\newblock \showarticletitle{$\{$GAZELLE$\}$: A low latency framework for secure
  neural network inference}. In \bibinfo{booktitle}{\emph{27th $\{$USENIX$\}$
  Security Symposium ($\{$USENIX$\}$ Security 18)}}.
  \bibinfo{pages}{1651--1669}.
\newblock


\bibitem[\protect\citeauthoryear{Kone{\v{c}}n{\`y}, McMahan, Ramage, and
  Richt{\'a}rik}{Kone{\v{c}}n{\`y} et~al\mbox{.}}{2016a}]%
        {konevcny2016federated1}
\bibfield{author}{\bibinfo{person}{Jakub Kone{\v{c}}n{\`y}},
  \bibinfo{person}{H~Brendan McMahan}, \bibinfo{person}{Daniel Ramage}, {and}
  \bibinfo{person}{Peter Richt{\'a}rik}.} \bibinfo{year}{2016}\natexlab{a}.
\newblock \showarticletitle{Federated optimization: Distributed machine
  learning for on-device intelligence}.
\newblock \bibinfo{journal}{\emph{arXiv preprint arXiv:1610.02527}}
  (\bibinfo{year}{2016}).
\newblock


\bibitem[\protect\citeauthoryear{Kone{\v{c}}n{\`y}, McMahan, Yu, Richt{\'a}rik,
  Suresh, and Bacon}{Kone{\v{c}}n{\`y} et~al\mbox{.}}{2016b}]%
        {konevcny2016federated2}
\bibfield{author}{\bibinfo{person}{Jakub Kone{\v{c}}n{\`y}},
  \bibinfo{person}{H~Brendan McMahan}, \bibinfo{person}{Felix~X Yu},
  \bibinfo{person}{Peter Richt{\'a}rik}, \bibinfo{person}{Ananda~Theertha
  Suresh}, {and} \bibinfo{person}{Dave Bacon}.}
  \bibinfo{year}{2016}\natexlab{b}.
\newblock \showarticletitle{Federated learning: Strategies for improving
  communication efficiency}.
\newblock \bibinfo{journal}{\emph{arXiv preprint arXiv:1610.05492}}
  (\bibinfo{year}{2016}).
\newblock


\bibitem[\protect\citeauthoryear{McKeen, Alexandrovich, Berenzon, Rozas, Shafi,
  Shanbhogue, and Savagaonkar}{McKeen et~al\mbox{.}}{2013}]%
        {mckeen2013innovative}
\bibfield{author}{\bibinfo{person}{Frank McKeen}, \bibinfo{person}{Ilya
  Alexandrovich}, \bibinfo{person}{Alex Berenzon}, \bibinfo{person}{Carlos~V
  Rozas}, \bibinfo{person}{Hisham Shafi}, \bibinfo{person}{Vedvyas Shanbhogue},
  {and} \bibinfo{person}{Uday~R Savagaonkar}.} \bibinfo{year}{2013}\natexlab{}.
\newblock \showarticletitle{Innovative instructions and software model for
  isolated execution.}
\newblock \bibinfo{journal}{\emph{Hasp@ isca}} \bibinfo{volume}{10},
  \bibinfo{number}{1} (\bibinfo{year}{2013}).
\newblock


\bibitem[\protect\citeauthoryear{Mishra, Rathee, Duong, and Yasuda}{Mishra
  et~al\mbox{.}}{2018}]%
        {mishra2018fast}
\bibfield{author}{\bibinfo{person}{Pradeep~Kumar Mishra},
  \bibinfo{person}{Deevashwer Rathee}, \bibinfo{person}{Dung~Hoang Duong},
  {and} \bibinfo{person}{Masaya Yasuda}.} \bibinfo{year}{2018}\natexlab{}.
\newblock \showarticletitle{Fast Secure Matrix Multiplications over Ring-Based
  Homomorphic Encryption.}
\newblock \bibinfo{journal}{\emph{IACR Cryptology ePrint Archive}}
  \bibinfo{volume}{2018} (\bibinfo{year}{2018}), \bibinfo{pages}{663}.
\newblock


\bibitem[\protect\citeauthoryear{Paillier}{Paillier}{1999}]%
        {paillier1999public}
\bibfield{author}{\bibinfo{person}{Pascal Paillier}.}
  \bibinfo{year}{1999}\natexlab{}.
\newblock \showarticletitle{Public-key cryptosystems based on composite degree
  residuosity classes}. In \bibinfo{booktitle}{\emph{International conference
  on the theory and applications of cryptographic techniques}}. Springer,
  \bibinfo{pages}{223--238}.
\newblock


\bibitem[\protect\citeauthoryear{Regulation}{Regulation}{2016}]%
        {regulation2016regulation}
\bibfield{author}{\bibinfo{person}{Protection Regulation}.}
  \bibinfo{year}{2016}\natexlab{}.
\newblock \showarticletitle{Regulation (EU) 2016/679 of the European Parliament
  and of the Council}.
\newblock \bibinfo{journal}{\emph{REGULATION (EU)}}  \bibinfo{volume}{679}
  (\bibinfo{year}{2016}), \bibinfo{pages}{2016}.
\newblock


\bibitem[\protect\citeauthoryear{Tourky, ElKawkagy, and Keshk}{Tourky
  et~al\mbox{.}}{2016}]%
        {tourky2016homomorphic}
\bibfield{author}{\bibinfo{person}{Dalia Tourky}, \bibinfo{person}{Mohamed
  ElKawkagy}, {and} \bibinfo{person}{Arabi Keshk}.}
  \bibinfo{year}{2016}\natexlab{}.
\newblock \showarticletitle{Homomorphic encryption the “holy grail” of
  cryptography}. In \bibinfo{booktitle}{\emph{2016 2nd IEEE International
  Conference on Computer and Communications (ICCC)}}. IEEE,
  \bibinfo{pages}{196--201}.
\newblock


\bibitem[\protect\citeauthoryear{Van~Dijk, Gentry, Halevi, and
  Vaikuntanathan}{Van~Dijk et~al\mbox{.}}{2010}]%
        {van2010fully}
\bibfield{author}{\bibinfo{person}{Marten Van~Dijk}, \bibinfo{person}{Craig
  Gentry}, \bibinfo{person}{Shai Halevi}, {and} \bibinfo{person}{Vinod
  Vaikuntanathan}.} \bibinfo{year}{2010}\natexlab{}.
\newblock \showarticletitle{Fully homomorphic encryption over the integers}. In
  \bibinfo{booktitle}{\emph{Annual International Conference on the Theory and
  Applications of Cryptographic Techniques}}. Springer,
  \bibinfo{pages}{24--43}.
\newblock


\bibitem[\protect\citeauthoryear{Yang, Liu, Chen, and Tong}{Yang
  et~al\mbox{.}}{2019}]%
        {yang2019federated}
\bibfield{author}{\bibinfo{person}{Qiang Yang}, \bibinfo{person}{Yang Liu},
  \bibinfo{person}{Tianjian Chen}, {and} \bibinfo{person}{Yongxin Tong}.}
  \bibinfo{year}{2019}\natexlab{}.
\newblock \showarticletitle{Federated machine learning: Concept and
  applications}.
\newblock \bibinfo{journal}{\emph{ACM Transactions on Intelligent Systems and
  Technology (TIST)}} \bibinfo{volume}{10}, \bibinfo{number}{2}
  (\bibinfo{year}{2019}), \bibinfo{pages}{1--19}.
\newblock


\bibitem[\protect\citeauthoryear{Yao}{Yao}{1986}]%
        {yao1986generate}
\bibfield{author}{\bibinfo{person}{Andrew Chi-Chih Yao}.}
  \bibinfo{year}{1986}\natexlab{}.
\newblock \showarticletitle{How to generate and exchange secrets}. In
  \bibinfo{booktitle}{\emph{27th Annual Symposium on Foundations of Computer
  Science (sfcs 1986)}}. IEEE, \bibinfo{pages}{162--167}.
\newblock


\bibitem[\protect\citeauthoryear{Zhu, Han, Mao, and Dally}{Zhu
  et~al\mbox{.}}{2016}]%
        {zhu2016trained}
\bibfield{author}{\bibinfo{person}{Chenzhuo Zhu}, \bibinfo{person}{Song Han},
  \bibinfo{person}{Huizi Mao}, {and} \bibinfo{person}{William~J Dally}.}
  \bibinfo{year}{2016}\natexlab{}.
\newblock \showarticletitle{Trained ternary quantization}.
\newblock \bibinfo{journal}{\emph{arXiv preprint arXiv:1612.01064}}
  (\bibinfo{year}{2016}).
\newblock


\end{thebibliography}

\end{document}